# Development of cooling system of solid state target for irradiation under proton beam of C18 cyclotron

***A.Avetisyan, R.Dallakyan, N.Dobrovolski, A.Manukyan, A.Melkonyan, I.Sinenko***

A. I. Alikhanyan National Science Laboratory (Yerevan Physics Institute) Foundation,
Yerevan, Armenia

## Abstract

In recent years, the possibility of direct production of the $^{99m}$Tc isotope (bypassing the parent $^{99}$Mo stage) for medical purposes using nuclear reactions on charged particle beams has been actively discussed around the world [1,2]. At A.I. Alikhanyan National Science Laboratory (Yerevan Physics Institute), an activity is underway to develop a technology for producing the $^{99m}$Tc isotope by irradiating a molybdenum target of $^{100}$Mo, pressed into a titanium base, with a proton beam of a C18 cyclotron [3,4,5].

One of the limitations of this technique is the utilization of heat released in the target as a result of proton energy loss [6,7].

The task of this work was the improvement of the thermal regime of a standard target for the C18 cyclotron due to a series of parallel grooves. Heat transfer experiments with prototypes of targets were carried out on a specially made test bench providing water cooling of its back side. A special Plexiglas-thermal block was made to study the thermal processes in the target.

The measurement results show that the above mentioned technique of processing the base of the target leads to a significant increase in the rate of cooling of the target, which will allow to irradiate at significantly higher proton beam intensities, which in its turn will increase the irradiation efficiency and reduce the cost of the final product.

## Introduction

The standard target base for Nirta Solid Compact (Compact Solid Target Irradiation System) is a titanium disk with a diameter of 24 mm and a thickness of about 2 mm. On one side, the disk has a cylindrical recess in the center with a diameter of 12 mm and a depth of about 1 mm as shown on Fig.1. Molybdenum powder is pressed into this recess.

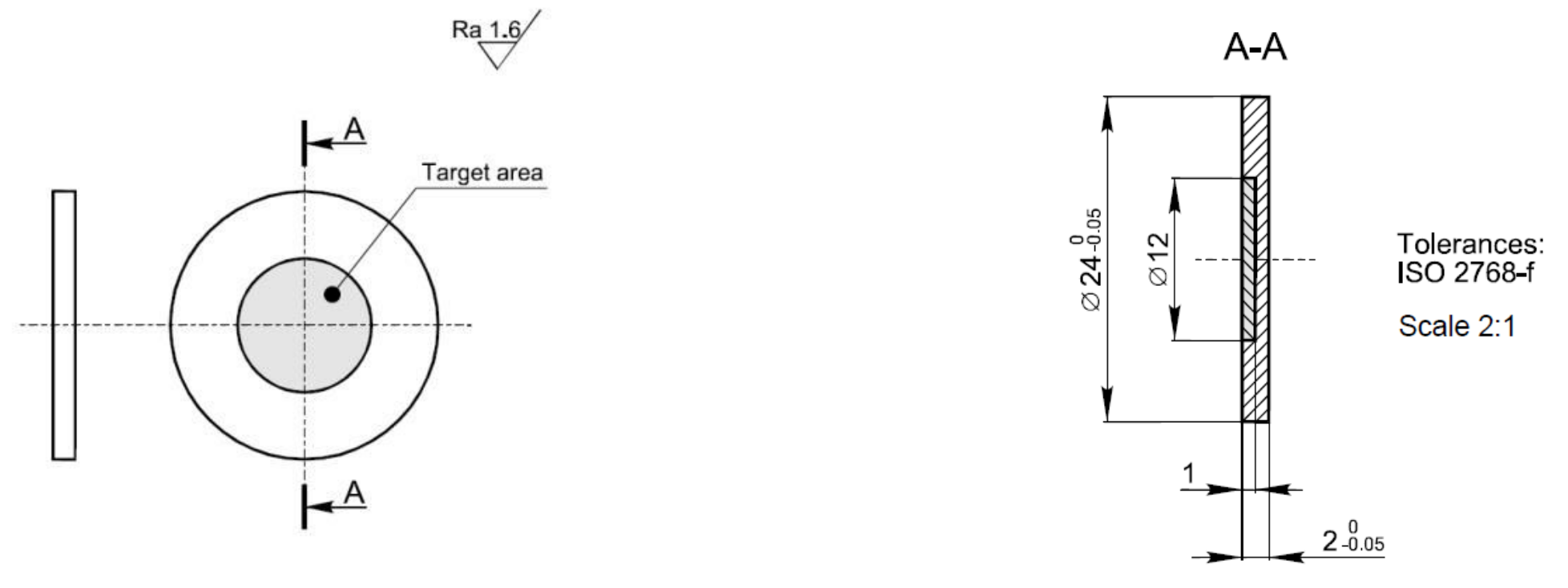

Fig. 1. Titanium target base.

In the standard target module the proton beam falls on the front side of the target, which is cooled by a stream of helium gas, and the back side of the target is cooled by running water under a pressure of 8 bars.

Usually the back of the target is a smooth surface. Modified target variants for cyclotrons with a ribbed back side are also described in order to improve the thermal contact of the target with cooling water [6-14].

The objective of this work was to improve the thermal regime of a standard target due to a series of parallel grooves engraved on the back of the target and oriented along the water flow, while maintaining overall dimensions. To study the effect of engraving on the heat transfer of the target, several prototypes of targets were made with various types of engraving:

1) for a smooth target without engraving (**S**mooth-**R**eference-**T**arget, SRT), which sets a reference point when analyzing the results

2) for a target with thin laser engraving (**L**aser-**E**ngraved-**T**arget, LET),

3) for targets with mechanically milled large grooves (**M**echanically-**M**illed-**T**arget, MMT) and

4) for a target with laser engraving, in which the residual roughness in the grooves is mechanically cleaned by hand - the target was obtained with both laser engraving and “milling” (**L**aser-**M**illed-**T**arget, LMT).

The additional water-cooling area for all target prototypes with engraving was approximately equal to the area of a smooth target without engraving. Thus that engraving doubled the surface which is cooled by water. This was enough to notice and measure the effect caused by the engraving.

Heat transfer experiments were done using special test bench provided water cooling of back side. The front side of the target prototypes was heated by a stream of hot air from an air gun with a controlled flow temperature in the range from room temperature to about 500° C. The measured value was the overheating of the target, i.e., the increase in its temperature versus the initial state, as a function of air temperature with various combinations of heating and cooling modes.

Experiments have shown that all types of engraving improve target cooling, but to a different degree. The greatest effect was obtained on a target with large milled grooves, the smallest - on a target with thin laser engraving.

## Description of targets and measuring setup

To accomplish heat transfer studies four above mentioned prototypes of the original target were made – SRT, LET, MMT and LMT. The main difference between all target prototypes and the original target was that the prototypes had composite structure shown on Fig. 2. Namely, every target prototype consisted of two titanium disks adpressed against each other. One disk was thin and smooth on both sides. It was used as a heat receiver. The receiver had a diameter of 24 mm, a thickness of 0.5 mm, and was located on the front side of the target.

The second, thicker disk formed the back of the target and served as a heat exchanger. Engraving was applied to the outward-facing side of the heat exchanger, and a copper insert with a built-in thermocouple junction for measuring temperature was embedded in the cavity on its opposite side.

The reference junction of the thermocouple was brought into thermal contact with a massive metal frame on which the entire installation was mounted, including the target and the tank with cooling water. The frame kept stable temperature during experiment and in essence served as a thermostat.

The thermocouple junction built into a copper insert to the target prototype made it possible to measure the temperature difference between the frame (thermostat) and a thin isothermal copper layer inside the target located at the same distance from both of its surfaces and conditionally referred to as the control temperature cross section. In the absence of external heating - a disconnected air gun - all elements of the device came into thermal equilibrium with each other and with a thermostat, and the thermocouple readings came to zero.

During the experiment, the front side of the target was blown with heated air with temperatures up to 500°C, and the back side was washed with a stream of water having the temperature of the one in thermostat.

Under such conditions, a temperature gradient was established inside the target, such a concept of “target temperature” lost its physical meaning and the only unambiguously determined value associated with the target temperature was the temperature difference indicated above between the frame (thermostat) and the control section. Let’s call this quantity “target overheating”. This definition is correct from a physical point of view and is intuitive, because with the air gun turned off, the “target overheating” defined by the above mentioned method is naturally zero, and when it is turned on, the control cross section heats up and the “target overheating” value quantifies this fact.

Schematically, the design of a target prototype is shown in Fig. 2. The left side of the figure shows a view on the target heat exchanger from the side of the heat receiver, but with the receiver removed. In this projection, the unmoved receiver would cover unimportant structural elements. For clarity, the left side of the figure is a photograph of one of the real target prototypes, and the right one is its schematic drawing with two cross sections.

The central cross section of target prototype with a vertical plane is shown on the right side of the schematic drawing, and the central cross section of the target prototype with a horizontal plane is shown below. To facilitate understanding of the relationship between the parts of the target, a receiver is also displayed on both cross sections. From this perspective, it does not cover up any structural details.

Both parts of the target prototype, receiver 2.1 and heat exchanger 2.2, are touched to each other through a thin layer of thermal paste 2.3 so that the side of the target with engraving 2.4 was facing outward, and thermocouple junction 2.5 was inside the target and had direct thermal contact with the disk through the copper insert 2.6 and thermal paste -heat exchanger 2.2.

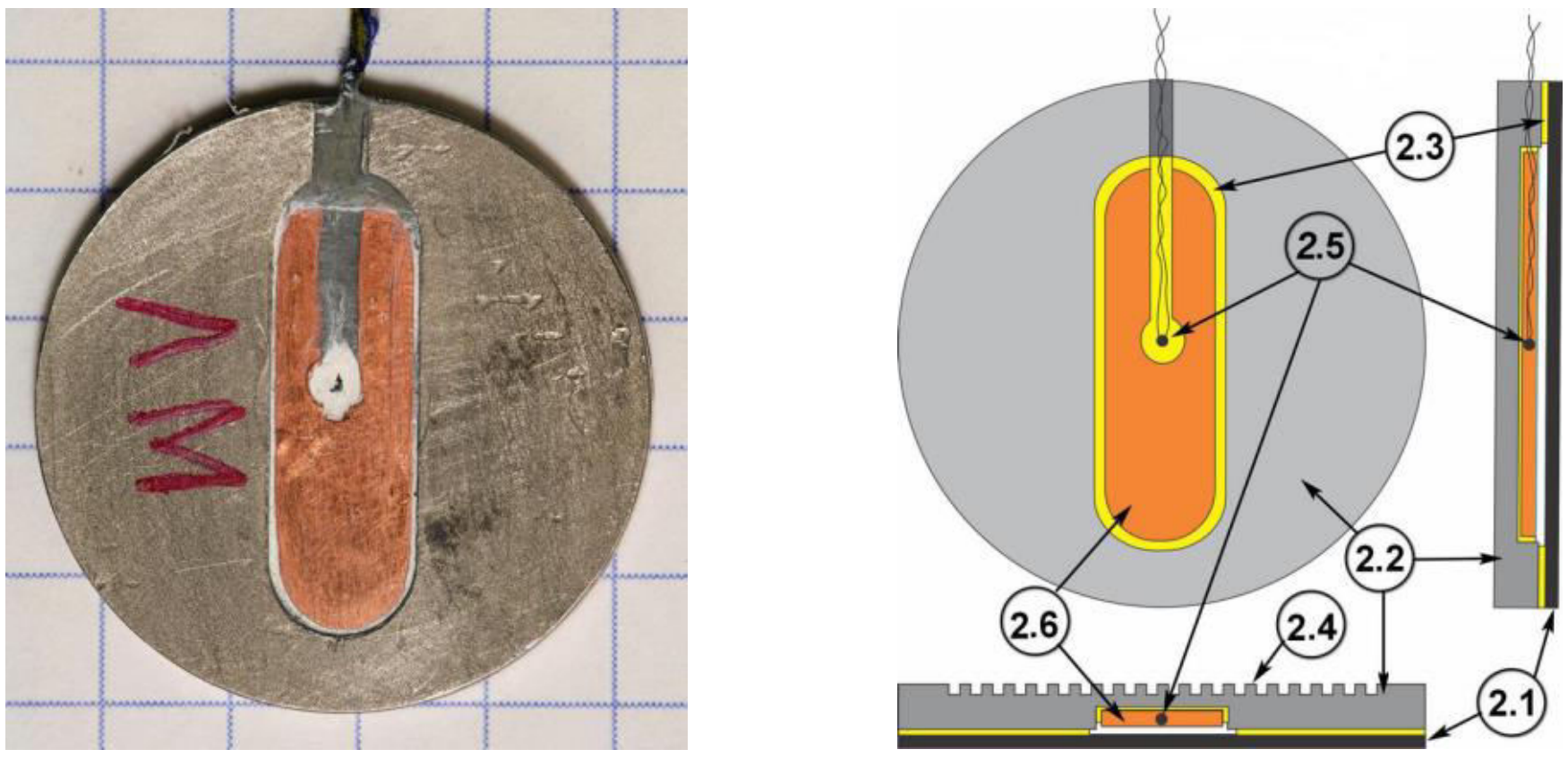

*Fig. 2.Schematic design of the target prototype.2.1 - Thin titanium disk, the front side of the target, a heat receiver. 2.2 - Thick titanium disk, heat exchanger.2.3 - Thermal grease, thermal-transfer medium. 2.4. – Engraving on the back of the target. 2.5 – Thermocouple junction with wires. 2.6 – Copper insert with a slit for mounting a thermocouple junction and wires*

Fig. 3 shows photographs of the backs of all four studied target prototypes. To better visualize the difference in the style of the engraving, we note that 50 equally spaced grooves were made on the LET and LMT targets with a laser, with the rib width, groove width, and its depth being approximately 0.2 mm each. On the MMT target, 20 equally spaced grooves were mechanically milled, the width and depth of the grooves, as well as the width of the edges, were also equal to each other but were larger and were approximately 0.5 mm.

On the zoomed fragment of the view of LET target, the specific character of the groove roughness is clearly visible – groove bottom is completely covered by rather deep annular craters of approximately the same size with a diameter of about 0.1 mm. Craters formed in places where the laser beam fell.

In order to assess the effect of craters on heat transfer from the target to the water, along with the LET target, another target was produced - LMT, with the same laser engraving, but in which the craters were mechanically smoothed. The degree of grooving of the grooves on the LMT target achieved in this way is illustrated by its zoomed view of fragment in Fig. 3, in which, along with the cleared grooves, three grooves are shown, which were left for comparison without stripping. They are located at the top of the enlarged fragment.

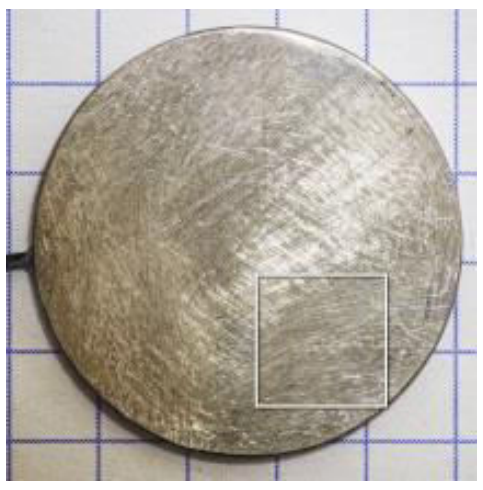

Back side of reference target prototype with smooth surface(SRT).

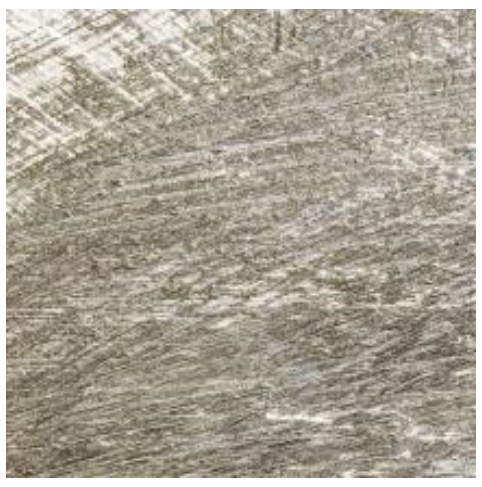

Zoomed view of smooth surface on indicated fragment of SRT.

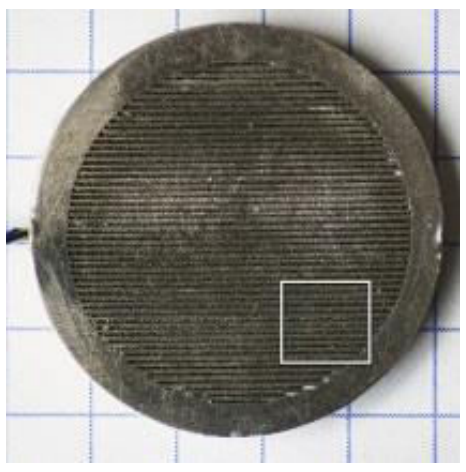

Back side of target prototype with thin laser engraving (LET).

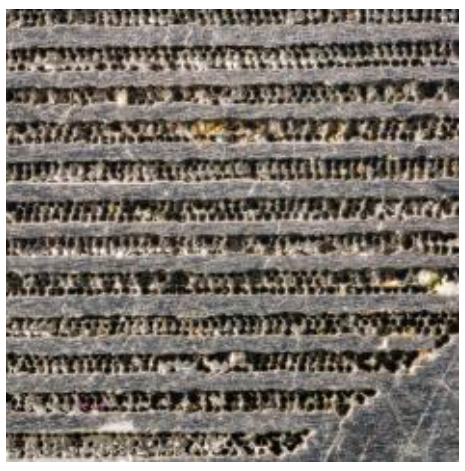

Zoomed view of grooves with craters on indicated fragment of LET.

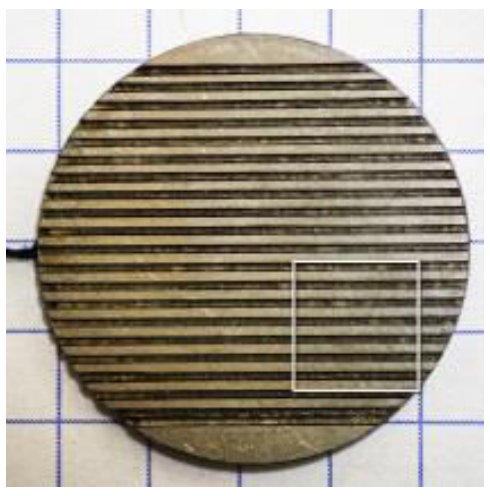

Back side of mechanically milled target prototype (MMT).

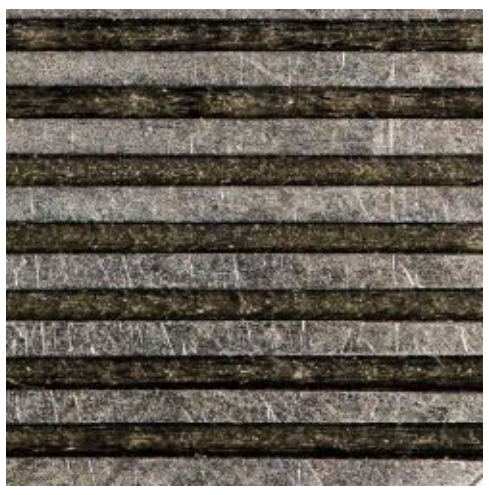

Zoomed view of smooth grooves on indicated fragment of MMT.

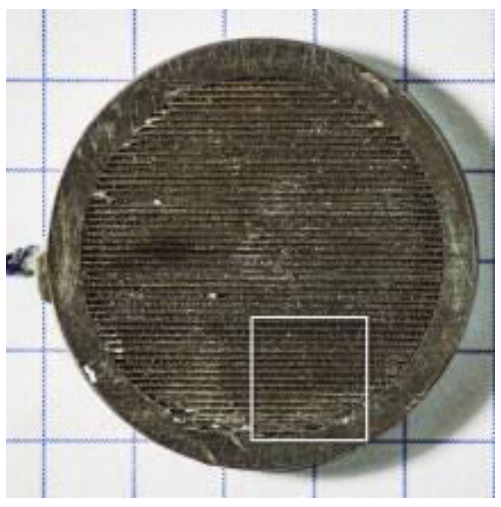

Back Side of TargetPrototype with laser engraving and mechanically smoothed cratersin grooves(LMT)

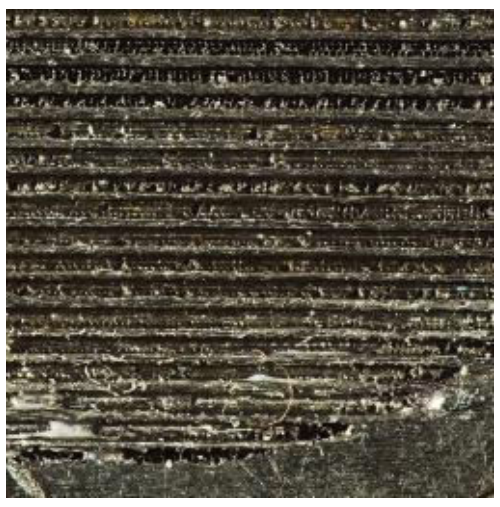

Zoomed view of grooves with mechanically smoothed craters on indicated fragment of LMT.

*Fig. 3.Photographs of the backs of the four studied targets. To the right of the overview photographs enlarged fragments of the lower right corners of the targets are shown. The location of each fragment in the photograph is approximately marked with a white square.*

The Figure 4 shows photographs of the back sides of the MMT and LEM targets in a view that sufficiently clearly shows the similarities and differences in the styles of their engraving.

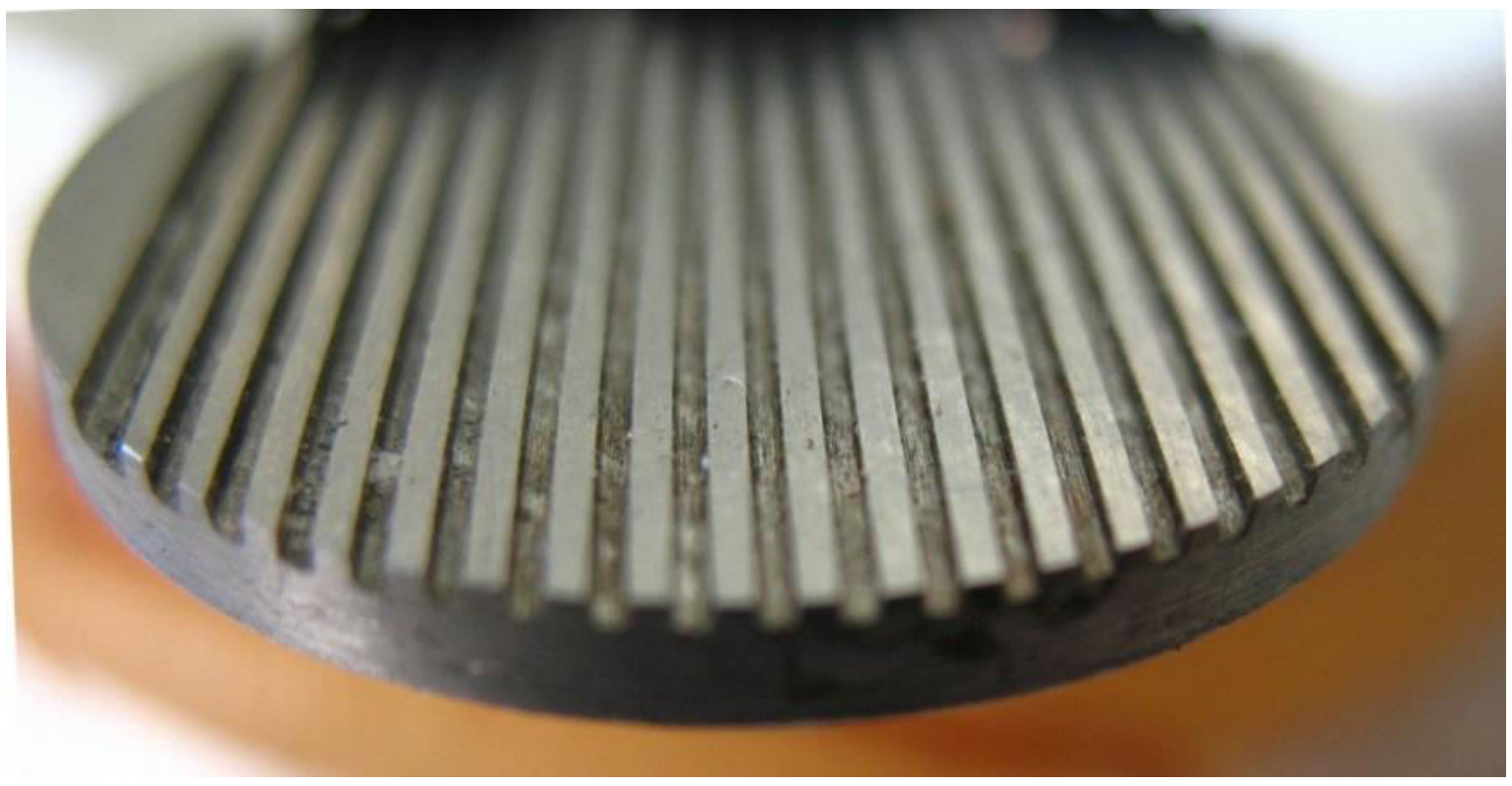

Smooth grooves on back side of mechanically milled target prototype (MMT).

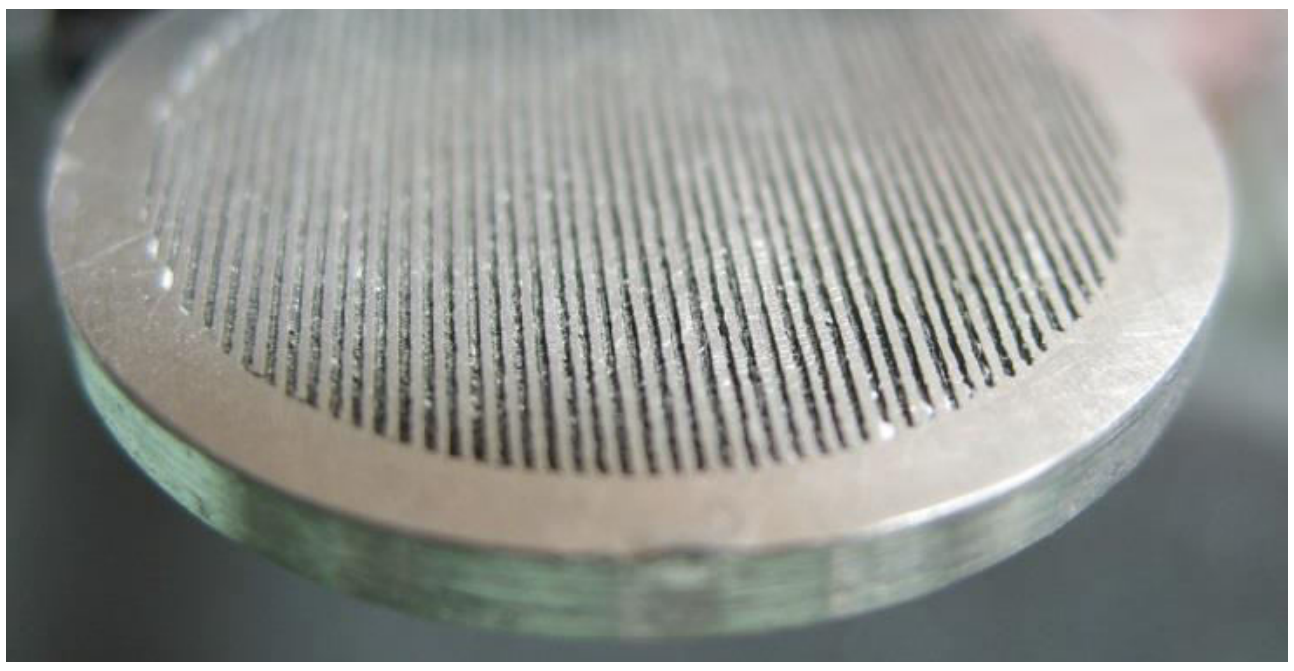

Grooves with craters on back side of target prototype with thin laser engraving (LET).

*Fig. 4.Photographs of the back sides of the LET and MMT targets at approximately the same scales and angles. The photographs clearly show that the grooves on both targets have an approximately rectangular cross section, but differ in size and quantity. On the MMT target, the grooves are larger and have a smooth surface. On the LET target, craters on the bottom of the grooves are visible even on a reduced scale. More clearly these craters are visible in fig. 3.*

The targets collected by the method described above were mounted in a Plexiglas thermal block. Plexiglas was chosen as the material for the thermal block due to its relatively low thermal conductivity. Experience has shown that parasitic heat transfer from the target receiver to cooling water through Plexiglas could be neglected compared to direct heat transfer from the receiver to water through the target material and the back of the target heat exchanger.

The thermal block consisted of two parts, namely a base and a cover. A photograph of the massive part of the thermal block- the base, is shown in Fig. 5. All the most important structural elements of the thermal block were concentrated just in the base.

The relative position of the main components of the measuring installation is shown in Fig. 6. The components themselves are listed in captions to the Fig. 5 and Fig. 6.

Cooling water inlet to the target was done through the base unit 1 along the pipe 2 and outlet along the second pipe 3. The water was circulated along an external closed circuit using a circulation pump developing a pressure of up to 1.5bar (this part of the device is not shown in the figure). The water flow and its pressure required for the experiment were set and regulated via bypass.

A hole was made in the center of the cover to supply hot air directly to the surface of the target receiver.

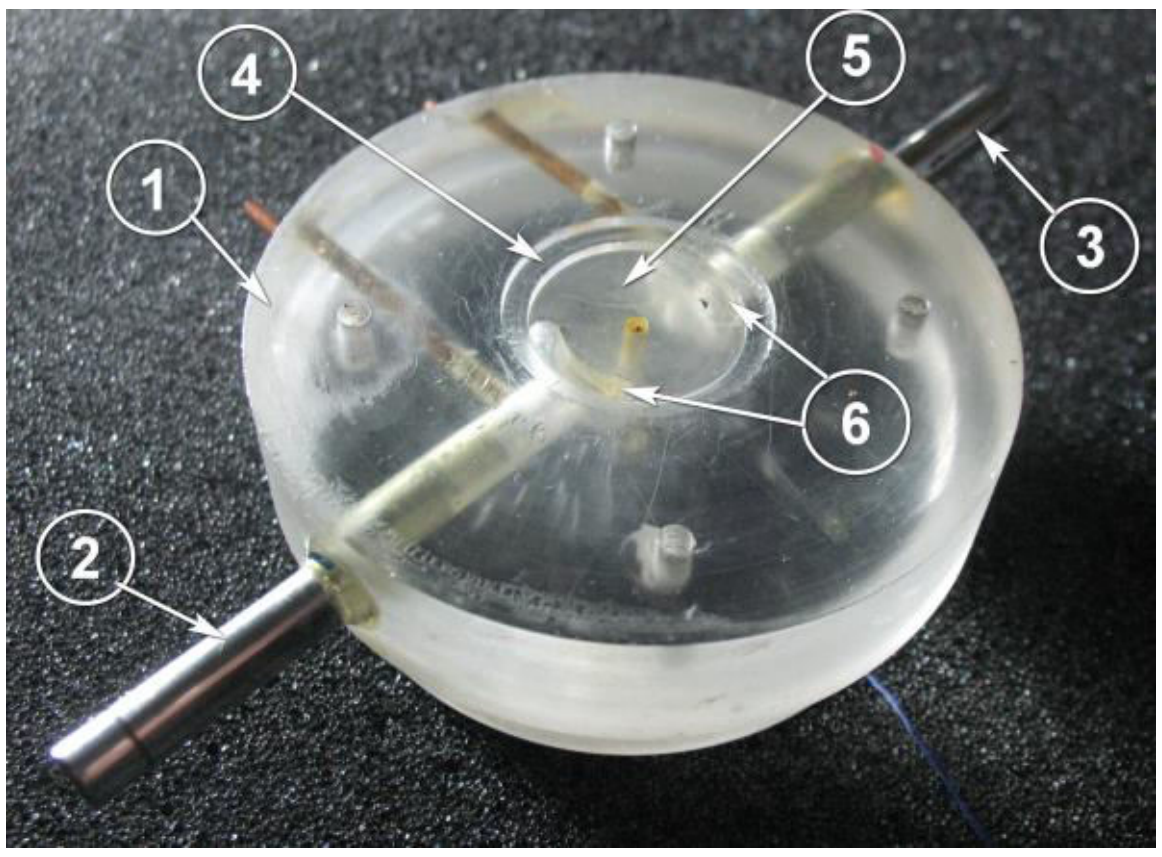

*Fig. 5.Plexiglas thermal block: 1 – the base of the thermal block. 2,3 - water cooling pipes. 4 - Seat for the target. 5 - Bottom of the heat exchange zone. 6 - Slotted guides of the water flow.*

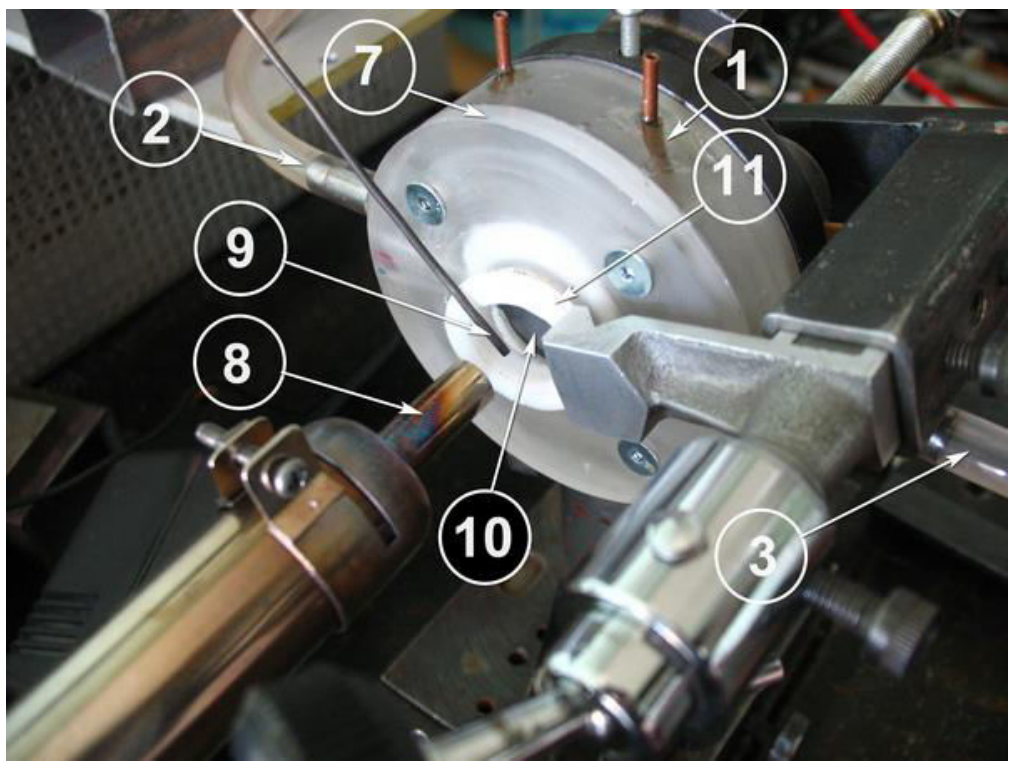

*Fig. 6.The main components of the measuring installation. Designations of common elements are continued from fig.5.1 – Plexiglas thermal block, a massive back part with nozzles - the base.2,3 – Water cooling pipes. 7 – The front part of a thermal blockwith a big hole in the center - its cover. 8 – The output nozzle of the air gun. 9 – Probe air temperature meter. 10 – The front side of the target prototype. 11– Teflon heat protective ring for Plexiglascover.*

The temperature of the air stream was set by the thermostat of the air gun and was measured independently using a separate industrial temperature meter with a digital display. This meter was equipped with a “K” type thermocouple sensor with a long probe, which allows measuring the temperature directly in front of the target’s face, which is clearly seen in Fig. 6.

Some schematic views of target prototypes and the thermal block assembly are shown in Fig. 7.

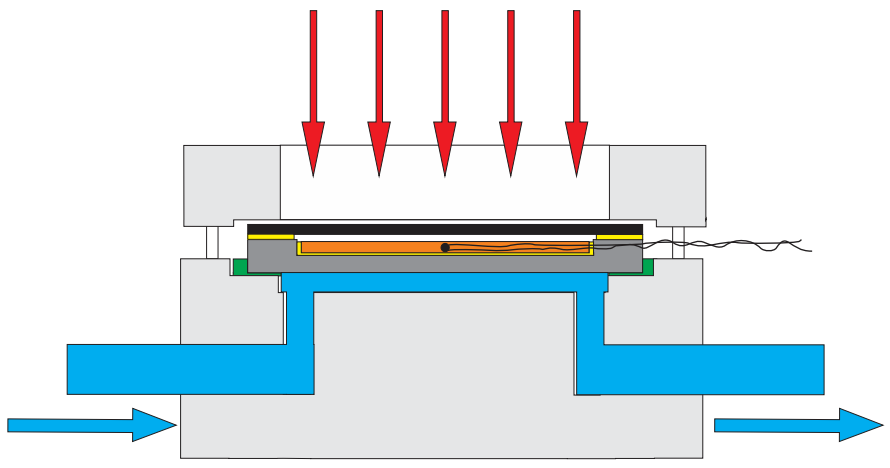

*Cross section of the thermal block with a target prototype along water stream, typical for all four types of target prototypes. Cooling water flows in the heat exchange zone from left to right along the engraved ribs, as shown by the arrows.*

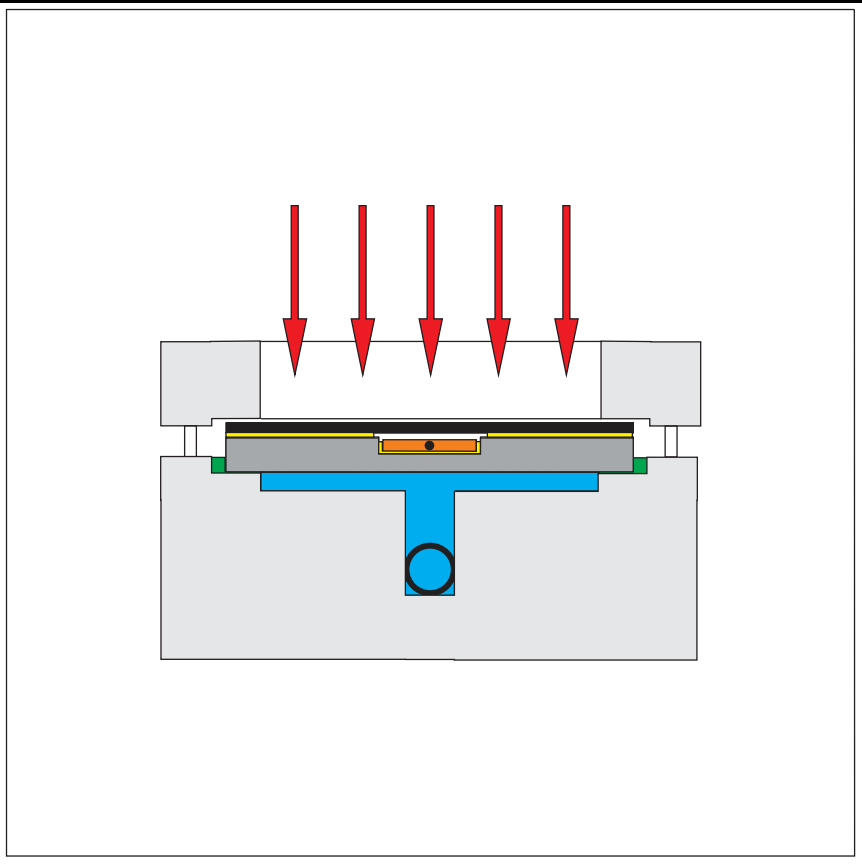

*Cross section of the thermal block and SRT target prototype across water stream. Cooling water in the heat exchange zone flows across the plane of the drawing.*

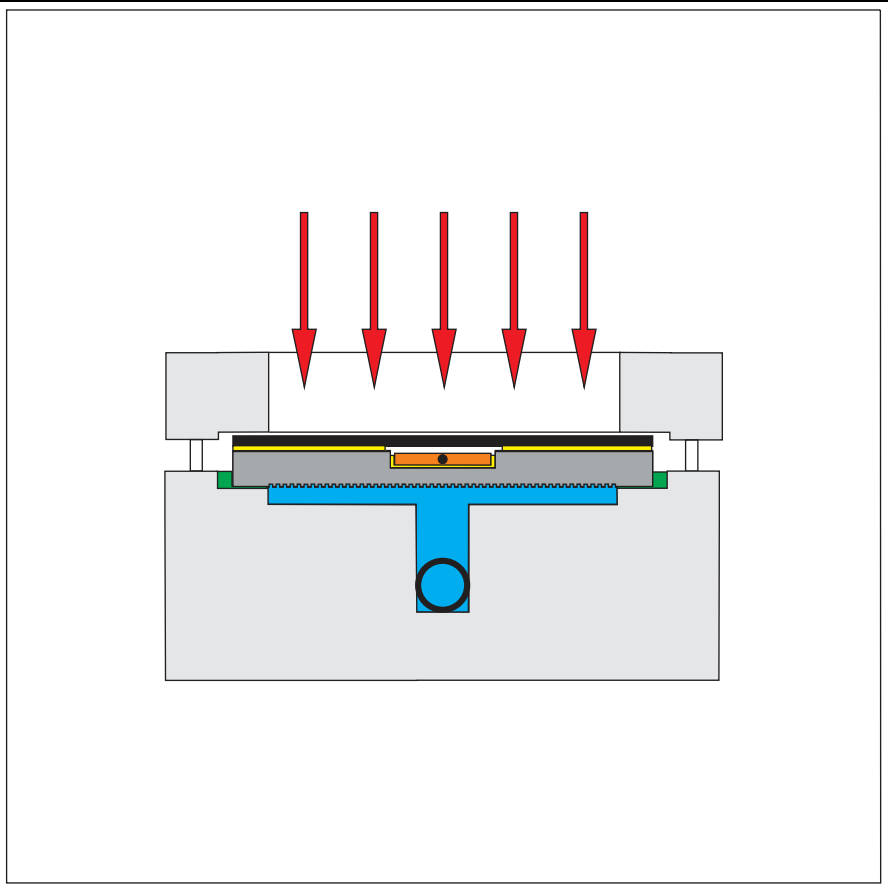

*Cross section of the thermal block and LET and LMT target prototypes across water stream. Cooling water in the heat exchange zone flows across the planeof the drawing.*

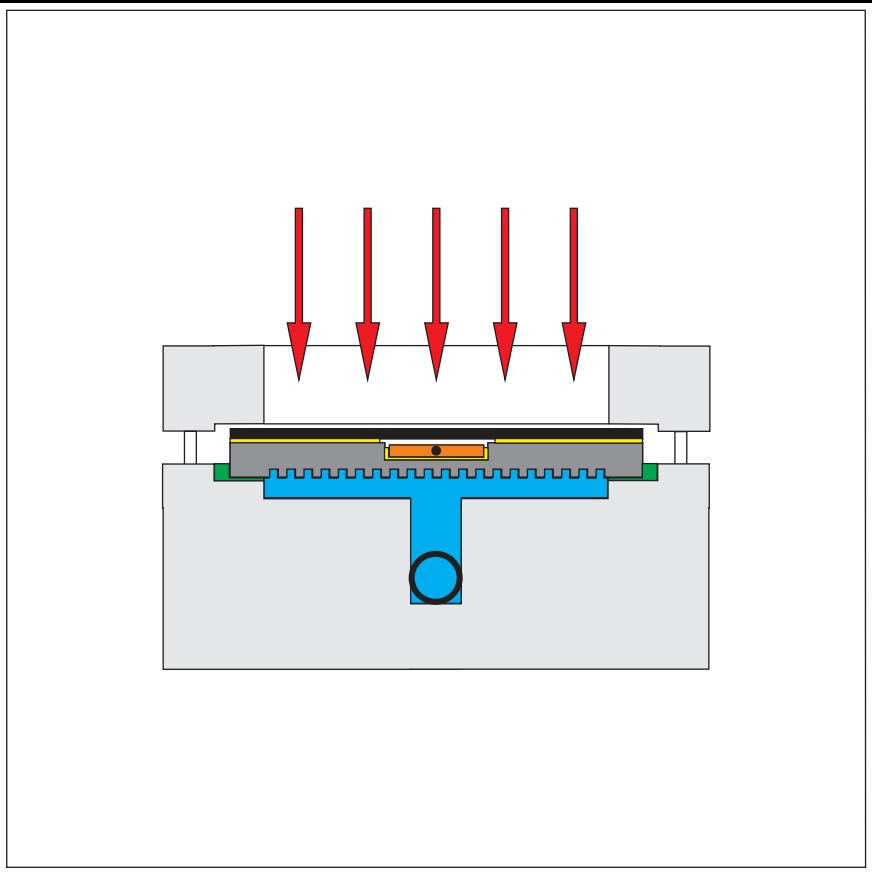

*Cross section of the thermal block and MMT target across water stream. Cooling water in the heat exchange zone flows across the plane of the drawing.*

*Fig.7.Schematic view on the measuring installation. Basic elements are designated bycolor:Light gray – Plexiglas thermal block base and cover. Dark gray – Thick titanium disk, a heat exchanger of target prototype. Black horizontal strip – Thin titanium disk, the front side of the target prototype, a heat receiver. Light brown – Copper insert with a slit for mounting a thermocouple junction and wires. Red – Stream of hot air. Blue – cooling water. Yellow –*

*Thermal grease, thermal-transfer medium between titanium disks. Green – Silicone sealant grease.*

## Measurement technique

Overheating of the target was measured with the help of a copper-constantan thermocouple manufactured by us, type "T". The thermoelectric signal from our thermocouple was measured with an "Aglient 34411A 6 ½ Digit Multimeter" digital multimeter with an accuracy of 1 μV.

The measurements were carried out under four different cooling modes, differing in the water flow rate, namely: 33 ml/sec (volume speed) (165 cm/sec – linear speed) - "fast water", 20 ml / sec (100 cm / sec) - "moderately fast water", 13 ml / sec (66 cm / sec) - "moderately slow water" and 9 ml / sec (45 cm / sec) - "slow water". In parentheses for each target, the calculated value of the linear velocity of the water flow in the heat exchange zone is given taking into account the actual dimensions of its passage section.

For definiteness, this work presents data only on the site of heating of the target, i.e. the dependence of the target overheating on the air flow temperature when the air temperature is unfolded from the room value; when the heating is off, to about 500° C - 520° C. The upper value of the heating temperature is limited on the one hand by the capabilities of the air gun, and on the other, by the heat resistance of the Plexiglas thermal block. The air flow rate in all series of measurements was fixed.

The sweep of the air temperature to the maximum value was carried out by the thermostat of the air gun and lasted about 300-400 seconds. At first, the heating rate was maximum, but in the region of 250 ° C it gradually decreased and continued to decrease as the temperature continued to increase, but already more slowly, and when it reached 530°C – 550°C, the speed almost reached zero. In fact, the temperature range up to 250 ° C was passed in the first 15-20 seconds, and the main heating time was in the remaining 250°C - 270°C.

## Results

The results are shown in graphs in Figures 7-10. The data are divided into four groups in accordance with the four speeds of cooling water, namely: for "fast water", for "moderately fast water", for "moderately slow water" and for "slow water". Each figure shows plots of the dependence of the target overheating on the temperature of the heating air, averaged over the results of three independent measurement sessions. To facilitate data analysis, charts are built on one scale.

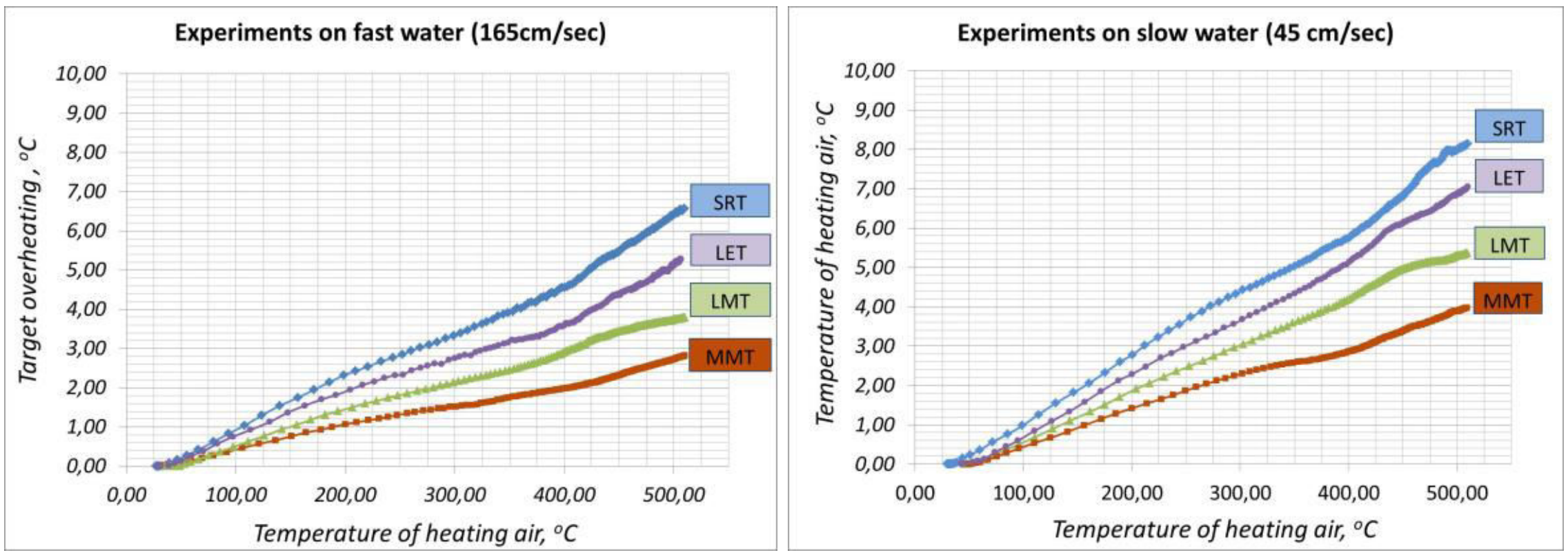

Fig. 7. Overheating of the target prototypes depending on the temperature of heating air flow in the cooling mode "fast water" at a speed of 165 cm/sec (left) and of 45 cm / sec(right).

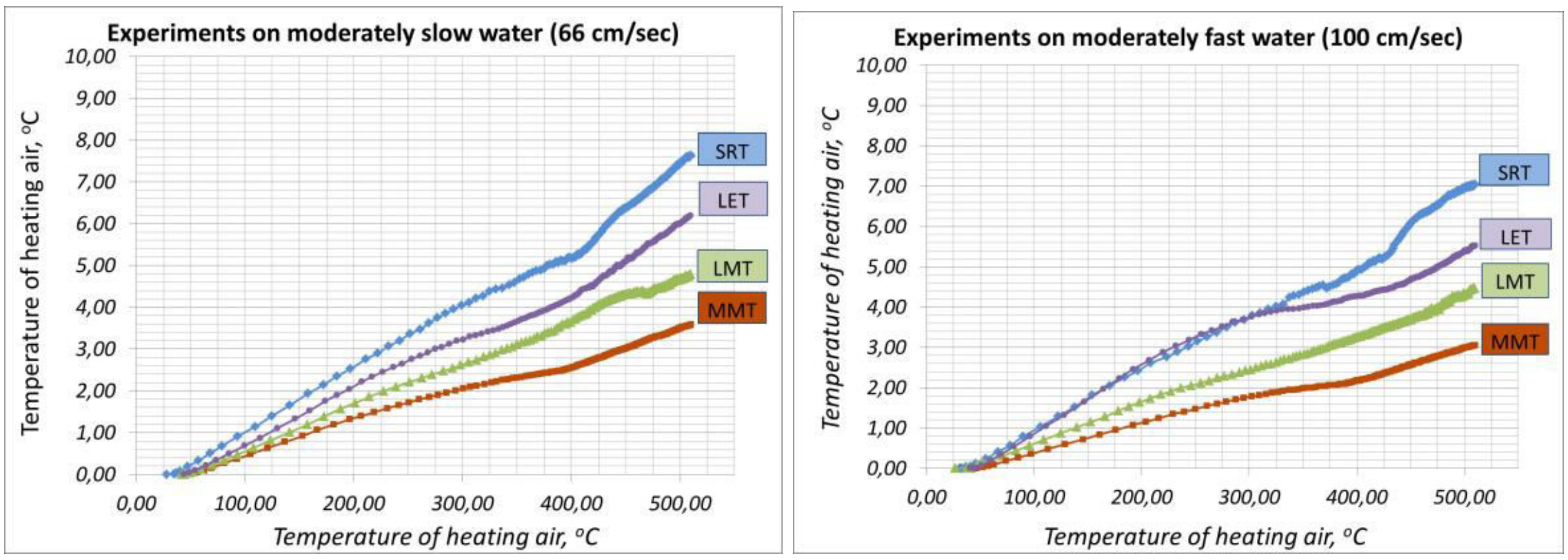

Fig. 8. Overheating of the target depending on the temperature of the air flow in the cooling mode "moderately slow water" at a speed of 66 cm / sec (left) and of 100 cm / sec (right).

# The discussion of the results

Table 1 lists the parameters related to the experiment — the overall characteristics of the targets, such as the number and dimensions of the grooves, as well as the area of the surface cooled by water. From the data given in parentheses in the fourth column, it can be seen that due to engraving, the cooled surface of the LET, MMT, and LMT targets was approximately doubled compared to a smooth SRT target. The last two columns show the numerical values of target overheating for the fastest and slowest water

The value of overheating is indicated both in absolute units - degrees Celsius, and in relative, as a percentage in relation to the values for a smooth target. Note that the lower the percentage, the lower the overheating of the target, and, consequently, the higher the efficiency of the target in terms of heat transfer.

Table 1.

| Target | Quantityof grooves | The size of a groove (edge) | The area of a cooled surface | Target overheat at temperature of heating air 500$^{O}$C on fast water | Target overheat at temperature of heating air 500$^{O}$C on slow water |
|---|---|---|---|---|---|
| | Piece | mm | mm$^2$ | $^{O}$C | $^{O}$C |
| SRT | 0 | 0 | 314(100%) | 6.4 (100%) | 8.0 (100%) |
| LET | 50 | 0.2x0.2 | 644 (205%) | 5.2 (81%) | 6.9 (86%) |
| LMT | 50 | 0.2x0.2 | 644 (205%) | 3.7 (58%) | 5.3 (66%) |
| MMT | 20 | 0.5x0.5 | 609 (194%) | 2.7 (42%) | 3.9 (49%) |

As can be seen from the table, there was a decrease in the overheating of the MMT target due to the deposition of grooves by about half compared with a smooth SRT target without engraving in proportion to the corresponding increase in the heat transfer area from the target to the water. Interestingly, the effect also depends on the speed of the cooling water - the effect on fast water is approximately 7% greater. It follows that in real experiments one should strive for the highest possible flow rates of cooling water.

This result was obtained on an MMT target with grooves about 0.5 mm deep.

It seemed that the specific roughness formed on the laser target with laser engraving due to technological mini-craters should increase the efficiency of its cooling with water. But the effect was the opposite. The fact that the reason for this lies precisely in the surface roughness, and not in the size and number of grooves, is proved by the increase in heat transfer on the LMT target during the cleaning of its grooves, because during the cleaning the number of grooves and their overall dimensions were preserved, only the roughness character was changed.

As can be seen from the last two columns of Table 1, the LMT target with stripped grooves in terms of cooling efficiency only slightly yielded to the MMT target with the smoothest grooves and the same additional area - 58% versus 42% in fast water and 66% against 49% in slow water. Recall that the lower the percentage, the lower the overheating of the target and, therefore, the higher the efficiency of the target in terms of heat transfer. We assume that the reason for the noted “lagging” of the LMT target is the fact that manual cleaning of the craters still did not give the degree of smoothness of the grooves that was obtained by machine milling in the manufacture of the MMT target.

The magnitude of the target overheating, in addition to the additional area in the grooves, also depends on the state of the cooling surface itself. Note that a smooth surface on targets with laser engraving can be obtained not only by mechanical cleaning of craters, but also by changing the laser mode accordingly. This part of the work remains to be done. The application of grooves on the back of the target by laser engraving is attractive because it gives much greater freedom in choosing the size of the groove, its shape and type of roughness.

**Conclusion**

A development of cooling system of solid state target for irradiation under proton beam of C18 cyclotron is discussed. The cooling thermal regime of C18 standard Nirta Solid Target was improved due to a series of parallel grooves engraved on the back of the target and oriented along the water flow. Several prototypes of targets were made with various types of engraving: a smooth target without engraving (**S**mooth-**R**eference-**T**arget, SRT), a target with thin laser engraving (0.2 mm x 0.2 mm grooves) (**L**aser-**E**ngraved-**T**arget, LET), a targets with mechanically milled large grooves (0.5 mm x 0.5 mm grooves) (**M**echanically-**M**illed-**T**arget, MMT) and target with laser engraving, in which the residual roughness in the grooves is mechanically cleaned by hand - the target was obtained with both laser engraving (0.2 mm x 0.2 mm grooves) and "milling" (**L**aser-**M**illed-**T**arget, LMT). The front side of the target prototypes was heated by a stream of hot air from an air gun with a controlled flow temperature in the range from room temperature to about 500° C. Experimetal measuring of the targets overheating has shown the highest cooling efficiency in case of MMT target despite it's a little small cooling area against LET and LMT targets.

**ACKNOWLEDGMENT**

This work was performed under the financial support of Armenian State scientific budget and IAEA CRP Contract 18029. The authors gratefully thank staff of Isotopes research and production department of A.Alikhanyan National Science Laboratory for their kind support.

**References.**